\documentclass[twocolumn]{aa}
\usepackage{graphicx}
\usepackage{txfonts}
\begin{document}
   \title{The outermost cluster of M31
	\thanks{Based on observations obtained with the Cassini Telescope
	        at the Loiano Observatory (Italy), operated by INAF - 
		Osservatorio Astronomico di Bologna.}
}
   \subtitle{}

   \author{S. Galleti\inst{1,2}, M. Bellazzini\inst{2}, L. Federici\inst{2},
          \and
 	   F. Fusi Pecci\inst{2}
          }

   \offprints{S. Galleti}

   \institute{Universit\`a di Bologna, Dipartimento di Astronomia
             Via Ranzani 1, 40127 Bologna, Italy\\
            \email{silvia.galleti2@unibo.it}
         \and
             INAF - Osservatorio Astronomico di Bologna,
              Via Ranzani 1, 40127 Bologna, Italy\\
              \email{
	      michele.bellazzini@bo.astro.it, luciana.federici@bo.astro.it, \\
	      flavio.fusipecci@bo.astro.it}             }
		
   \date{Received February 10, 2005; accepted March 1, 2005}

   \abstract{We report on the identification of a new cluster in the far halo
   of the M31
   galaxy. The cluster, named Bologna~514 (B514) has an integrated magnitude 
   $M_V=-8.5 \pm 0.6$, and a
   radial velocity, as estimated from two independent low-resolution spectra,
   $V_r=-456 \pm 23$ km/s, which fully confirms its membership to the M31
   system. The observed integrated spectrum is very similar to those of 
   classical globular clusters.
   Being located at $\simeq 4^o$ ($\simeq 55$ kpc in projected distance)
   from the center of the parent galaxy, B514 is by far the most remote M31
   cluster ever discovered. Its projected position, near the galaxy major axis,
   and M31-centric velocity, similar to that observed in the outermost regions
   of the HI rotation curve, may indicate that it belongs to the
   subsystem of M31 clusters that has been recently proposed (Morrison et al. \cite{morrison} )
   to be part of the dynamically-cold thin disc of the galaxy.  
   
   }

   \maketitle
%

\section{Introduction}

The globular cluster system of M31 (the Andromeda galaxy) has been the subject 
of deep investigations since the dawn of extragalactic astronomy  (Hubble
\cite{hubble}). M31 is the only giant spiral galaxy beyond the Milky Way that
we can hope to study in some detail, and globular clusters (GCs) are generally
recognized as  excellent tracers of the physical properties of the parent
stellar system (Hodge \cite{h_book}).  The comparative study of the GC systems
of M31 and of the Milky Way  may also provide very fruitful insights on the
formation history of the parent galaxies as well as important constraint to
theories of GC formation  (see Harris \cite{h01}, and references therein). 
Finally, the kinematics of GC provides a classical tool to estimate the mass
distribution of the parent galaxy out to large distances (see, Evans et al.
\cite{evans}, and references therein, for  the case of M31).

For the above reasons, the observational effort to find out M31 globulars and
to measure their fundamental parameters has been large and extensive in the
last 50 years (see, for example  Vete$\check{s}$nik \cite{vet62}, van den Bergh
\cite{syd67,syd69}, Baade \& Arp \cite{BA}, Sargent et al. \cite{sg77},
Crampton et al. \cite{ccsc85} Battistini et al. \cite{bat80,bat82,bat87,bat93},
Sharov et al. \cite{sha95}, Mochejska et al. \cite{Mochejska} and references
therein) and is still continuing (Barmby et al. \cite{barm00,barm01}, Perrett
et al. \cite{perr_cat}, Galleti et al. \cite{silvia}). The quest for clusters
located at large projected distances from the center of the galaxy is of
particular relevance since these object would allow new insight into unexplored
regions of M31 and  they would provide the maximum ``leverage'' to constrain
the mass  distribution of the galaxy (see, for example Federici et al.
\cite{luciana} and  Evans et al. \cite{evans}, and references therein). 

In this framework, we have recently started a programme aimed at the
identification of new candidate globular clusters and at the confirmation of
known candidates in the outermost regions of M31.  During a pilot observational
project we took low resolution spectra (${\Delta \lambda\over{\lambda}}\sim
1300$) of two promising candidates selected from the Extended Source Catalogue
of 2MASS  (XSC, Cutri et al. \cite{cutri}).  One of them turned out to be a
genuine M31 cluster, by far the most distant from the center of the galaxy that
has been ever identified. In the present contribution we briefly report on this
new finding. 

In the following we will adopt the  distance modulus of M31 from McConnachie et
al. (\cite{mcc}), $(m-M)_0=24.47$ (see also Durrell et al. \cite{durrell}), the
average color excess $E(B-V)=0.11$, from Hodge (\cite{h_book}), and the line of
sight systemic velocity of $V_r=-301$ km/s, from van den Bergh
(\cite{sydbook}).  All the parameters for known M31 globulars are from the
comprehensive compilation by Galleti et al. (\cite{silvia}) with the only
exception of radial velocities and spectroscopic metallicity estimates
that are drawn from Perrett et al.
(\cite{perr_cat}), when possible, and from  Huchra et al. (\cite{bh}) when
lacking in the former list.  All over the paper we will only consider GC
candidates whose nature has been confirmed by means of spectroscopy and/or high
resolution imaging (confirmed clusters, see Galleti et al. \cite{silvia}, and
references therein), if not otherwise stated.

\section{Selection of distant candidates}

In Galleti et al. (\cite{silvia}, hereafter G04) we provided J, H and K 
integrated magnitudes of
693 candidate M31 GCs obtained from the All Sky Data Release of 2MASS.  Of the
279 confirmed clusters included in this set, the large majority (258) were
found in the Point Source Catalogue (PSC) of the survey, while just 21 were
included in the XSC. This strongly suggest that if a M31 GC candidate is
included in the XSC it is a bona-fide extended source, i.e. it is not a
foreground stars\footnote{Foreground Galactic stars are the major source of
false M31 GC candidates, together with background galaxies.  However the nature
of the latter can be firmly established just on the basis  of their radial
velocity that shall be much higher than the systemic velocity of M31, while a
foreground stars cannot be always identified on the basis of its velocity
alone.}. Intrigued by this possibility we searched for new GC candidates among
XSC sources within a $\sim 9\degr \times 9\degr$ area centered on M31 (and
excluding the innermost $3\degr \times 3\degr$ field already studied in G04).
Using suitable combinations of the structural
parameters  provided in the XSC (ellipticity, radius, etc.,  see Galleti et al.
\cite{silviab} for some examples of the adopted techniques) we selected $49$
candidates worth of further investigation. DSS-2 images of these candidates
were visually inspected and 19 of them were rejected as obvious galaxies during
this phase. As a pilot project we decided to take spectra of the two most
promising candidates (as judged from their appearance on DSS-2 images), whose
2MASS names and main characteristics are reported in Table~1.

\begin{table*}
\caption{Observed parameters of the targets}             
\label{table:1}      
\scriptsize
\centering                          
\begin{tabular}{l c c c c c c c c}        
\hline\hline                 
2MASS ID&  $RA(J2000)$ & $DE(J2000)$ & r & J & H & K & $V_r$ [km/s] & $V_r$ [km/s]\\    
        & 		 &		&     &   &   &   &  Nov  & Jan \\    
\hline                        
2MASX-J00310983+3753596& $0^h31^m 9.8^s$ & $37^o 53\arcmin 59.6\arcsec$& $5.4\arcsec$& $14.272\pm0.048 $ &$ 13.607\pm0.059 $ &$13.661\pm0.087 $ &$-459 \pm 12$& $-453 \pm 29$ \\      
2MASX-J00262028+4400377& $0^h 26^m 20.3^s$ & $44^o 00\arcmin 37.7\arcsec$&$6.6\arcsec$&$14.122\pm0.042$ &$ 13.430\pm0.057$ &$13.184\pm 0.061 $&$20069 \pm 22$ &$20097 \pm 76^1$ \\      
                       &   & & & & & &$20058 \pm 21$ & 	     \\      
\hline                                   
\end{tabular}
{\begin{flushleft}{r is the Kron circular aperture radius in the J band 
(see Cutri et al.\cite{cutri} for details); J,H and K integrated magnitudes
 are derived as in Galleti et al. \cite{silvia}. $^1$ From a spectrum obtained 
 with a slit of $2\arcsec$, see text.}
\end{flushleft}}
\end{table*}

\subsection{Observations}

We obtained 5 low-resolution integrated-light spectra of the two candidates
globular clusters. The observations were performed in November 16-17, 2004 and 
January 02-05, 2005, with the BFOSC\footnote{See  {\tt 
http://www.bo.astro.it/~loiano/TechPage/pagine/
BfoscManualTechPage/BfoscManual.htm} for further details on  the instrument.}
spectrograph operated at the 1.52m Cassini Telescope of the Loiano Observatory,
near Bologna (Italy). The detector was a thinned, back illuminated EEV CCD,
with $1300 \times 1340 ~px^2$.  During the first night of the January run a
slit of $2\arcsec$ was used to match poor seeing conditions. A slit of
$1.5\arcsec$ was used  during the other nights. As a consequence, the
instrumental resolution was slightly lower during the first night of the
January run. The adopted set-up (slit width $1.5\arcsec$) provides a spectral
resolution $\Delta \lambda = 4.1 \AA$ over the range $4200 \AA < \lambda < 6600
\AA$. The pixel scale is $1.94 \AA~px^{-1}$. We took a He-Ar calibration lamp
spectrum after each scientific exposure, maintaining the same pointing of the
telescope.  In the November run we obtained two spectra of
2MASX-J00262028+4400377 and  one of 2MASX-J00310983+3753596; in the January run
we took one spectrum for  each object ( the spectrum of 2MASX-J00262028+4400377
was observed with the slit of  $2\arcsec$). The exposure time was 2700 s for
each target spectrum,  with a  characteristic signal-to-noise ratio $S/N \sim
25$.   During each observing night we also observed, with the same set-up, four
template targets that we adopted as radial velocity standards: the M31 globular
clusters B158 and B225 (heliocentric radial velocities from Perrett et al.
\cite{perr_cat}) and the two stars HD12029, HD23169 (heliocentric radial
velocities from the SIMBAD database).  In the first night of the January run
only B225 was observed as template  with the slit of $2\arcsec$. All the
template spectra have $S/N > 70$.

All the scientific spectra were processed with standard reduction techniques in
IRAF. In summary, after subtracting a masterbias, all frames were divided by  a
normalized flat-field image. Subsequently, the IRAF/twodspec procedures were
used to sky-subtract, extract and calibrate in wavelength all spectra. The
heliocentric radial velocities ($V_r$) of the candidates globular clusters 
were obtained by cross-correlation (CC) with the high $S/N$ spectra of the
templates, using the IRAF/fxcor package (see Tonry \& Davis  \cite{tonry} for
details of the technique).

The obtained solutions were  unambiguous in all cases, i.e., a well defined 
single peak of the CC function was found, with heights in the range  $0.3\le
CC\le 0.55$; the typical uncertainty on the single estimate  is $\sim 50$ km/s
(FWHM of the CC peak). For each of our target spectra we took the average of
the $V_r$ estimates from the different templates (except for the spectrum of 
2MASX-J00262028+4400377 taken in the January run, for which only  one template
spectrum was obtained with the $2\arcsec$ slit setup).  The standard deviations
of these estimates were summed in quadrature with the errors in the velocity
of  the templates to obtain the uncertainty on the $V_r$ from each single
spectrum.  The $V_r$ estimates obtained in this way from the spectra of the
same target  taken in different nights were fully compatible  among each other,
within the uncertainties. Hence we took the average of the  $V_r$ obtained from
the different spectra (nights) as our final estimate of the heliocentric radial
velocity. The final total uncertainty was obtained by  averaging the
uncertainties from the (two-three) independent estimates. As a consistency test
we derived the velocity of each template cluster by CC with the others
templates and in both case we obtained estimates within $\pm 2.0$ km/s of the
published values  (Perrett et al. \cite{perr_cat}). 

\begin{figure}
   \centering
   \includegraphics[width=8cm]{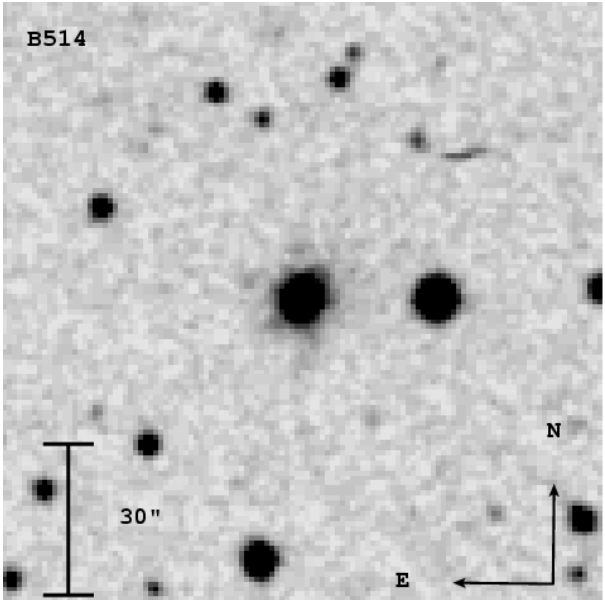}
      \caption{$2\arcmin \times 2\arcmin$ DSS-II(Red) image centered
      on B514 (http://archive.eso.org/dss/dss). Note the fuzzy nature
      of the B514 image with respect to the nearby bright star located
      $\sim 20\arcsec$ to the West of the cluster.
 	}
         \label{map}
   \end{figure}

\begin{figure}
   \centering
   \includegraphics[width=8cm]{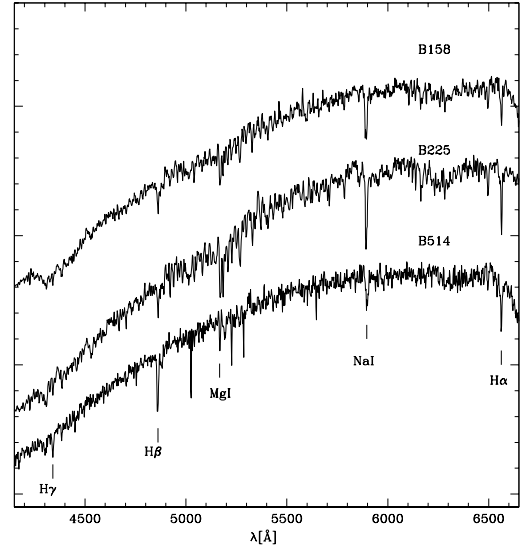}
      \caption{Spectra of the confirmed M31 globular clusters 
        B225, B158 compared with that of B514. The most relevant spectral 
	features are marked, as a reference. The ordinate axis is in arbitrary 
	units. The spectra are corrected for the 
	heliocentric radial velocity (e.g. shifted to rest-frame).	
              }
         \label{spectra}
   \end{figure}

The observed parameters of our targets are shown in Table~1.  The very high
velocity reveals that 2MASX-J00262028+4400377 is clearly a background galaxy at
$z\simeq 0.07$.  On the other hand 2MASX-J00310983+3753596 has a velocity
within $\sim 150$ km/s of the systemic velocity of M31 (the straight velocity
dispersion of M31 GCs is $\simeq 160$ km/s) hence it can be safely classified
as a confirmed M31 cluster. Following the naming convention by Battistini et
al. (\cite{bat87}) we christen the newly discovered cluster Bologna~514 (B514).
The object appears clearly non-stellar even in DSS-2 images (Figure~\ref{map})
and has a spectrum typical of known Andromeda GCs (see Figure~\ref{spectra}). 
The main derived parameters for B514 are reported in Table~3. No matching
counterpart has been found in the NED and SIMBAD databases, except for the
obvious 2MASS entry. In addition, other databases such as Vizier (Ochsenbein et
al. \cite{vizier})  were consulted. We found that B514 is listed in the
HYPERLEDA catalogue (Paturel et al. \cite{leda}) as an extended source detected
on the DSS (PGC2112471), lacking any information on the velocity.   

While this contribution was near to completion we realized that one of the new
M31 GC candidates found by Huxor et al. (\cite{hux}) has a position similar to
B514 (according to their Fig.~1, since they don't provide  the coordinates or
the radial velocity estimates for the presented candidates).  Upon direct
request, these authors kindly confirmed that their outermost candidate is
indeed the  same object we are dealing with in the present contribution 
(Huxor et al.,
private communication). Hence B514 has been independently identified  also by
this team. It is worth noting that in the image reported as Fig.~3 of Huxor et
al. (\cite{hux}), the object clearly appears as a partially  resolved stellar
system.

%
%
%
%
%
%
\section{Bologna~514: the most remote M31 cluster}

The 2MASS NIR integrated magnitudes of B514 indicates that it is a quite
luminous object, e.g. among the $\sim 40$\% brightest M31 GCs in K band  
in the extensive database by G04. 
As a reference for future observations it may be useful to provide
a guess of the visual integrated magnitude.
Lacking any visual photometry, 
we obtained a rough estimate of the V
integrated  magnitude by fitting the linear relations between V and J,H,K
magnitudes for the confirmed clusters of M31, 
taken from the G04 catalogue:
$$V=0.81J+4.70~(r.m.s=0.46~mag)$$
$$V=0.71H+6.62~(r.m.s=0.49~mag)$$
$$V=0.79K+5.50~(r.m.s=0.56~mag)$$ 
We obtained  $V=16.26\pm 0.46,
16.28\pm 0.50, 16.29 \pm 0.57$ from the V$vs.$J, V$vs.$H and  V$vs.$K
relations, respectively. As a final conservative estimate we take  $V=16.3 \pm
0.5$. With our distance and reddening assumptions this corresponds to an
absolute integrated magnitude $M_V=-8.5 \pm 0.6$, that places B514 among the
$\sim 30$\% most luminous clusters of M31 also in the V band (G04). 

The most remarkable characteristic of the clusters is its position within the
M31 system. B514 is located at $R\simeq 4^o$ from the center of the galaxy,
corresponding to a projected distance $R_p=54.7$ kpc, hence it is by far the
most distant M31 cluster presently known.  The previous ``record holder''
(G~1=Mayall II, see Meylan et al. \cite{g1}),  lies at $R\simeq 2.5^o$
($R_p=34.2$ kpc).  Most interestingly, the cluster lies just $\simeq 14
\arcmin$ apart of the galactic major axis, on the south-western ideal extension
of the disc (see Fig.~\ref{xy}). Moreover its ``M31-centric'' line-of-sight
velocity ($V_{M31}=-151 \pm 19$ km/s) places it around the extrapolation at
large  distance of the HI disc velocity curve (see Fig.~\ref{rotaz}). 

\subsection{Metallicity and age}

To have an indication of the average
metal content we used the relation between [Fe/H] and $(J-K)_0$ provided by
Barmby et al. (\cite{barm00}). Adopting $(J-K)_0=0.55\pm 0.10$ for B514 (see
Tab.~1) we obtain [Fe/H]=$-2.0 \pm 0.9$, suggesting that the cluster is quite
metal poor (at least if its age is similar to that of the other M31 GCs).
As a comparison, we used the same relation for the clusters G~1 and G~2, 
that are
also located at large distances from the center of M~31 and for which
spectroscopic estimates of the metallicity are available in the literature (see
Sect.~3.1 for further discussion about these clusters in relation with B514).
According to G04, G~1 has $(J-K)_0=0.75\pm 0.10$ from which we obtain 
[Fe/H]=$-0.8 \pm 0.9$, in good agreement with the estimates by Huchra, Brodie 
\& Kent (\cite{bh}) and Meylan et al. (\cite{g1}) who find 
[Fe/H]=$-1.08 \pm 0.09$ and [Fe/H]=$-0.95 \pm 0.09$, respectively; 
G~2 has $(J-K)_0=0.62\pm 0.10$, corresponding to [Fe/H]=$-1.6 \pm 0.9$, in good
agreement with [Fe/H]=$-1.70 \pm 0.36$ by Huchra, Brodie \& Kent \cite{bh}.

\subsubsection{Line indices}

We have measured the line indices $H_{\beta}$ and $Mgb$ from our spectra of M31
GCs (B158, B225 and B514), according to different definitions (Brodie \& Huchra
\cite{bh0}, Perret et al. \cite{perr_cat}, and Trager et al. \cite{trager}). 
The obtained
estimates in the scale by Brodie \& Huchra \cite{bh0} are reported in table 2
together with the values published by the same authors for B158 and B225. In all
the considered cases the agreement of our estimates with the measures 
available in the literature for these clusters
is very good. 
We choose the $H_{\beta}$ and $Mgb$ indices because (a) they are
quite sensitive to age and metallicity, respectively, (see, for example,
Buzzoni \cite{buz95}, and references therein) and (b) they
correspond to rather strong features in our spectra, thus they can be reliably
measured. 

Adopting the [Fe/H] $vs$ Mgb calibration by Perret et al. (\cite{perr_cat}) we
find [Fe/H]=$-1.85\pm 0.3$ for B514; with the analogous relation provided by
Brodie \& Huchra (\cite{bh0}) we obtain [Fe/H]=$-1.79\pm 0.3$. As a final
metallicity we take the average of the two [Fe/H]=$-1.8\pm 0.3$, in good
agreement with the photometric estimate derived above. 
The fraction of M31 GCs with [Fe/H]$\le -1.5$ among the confirmed clusters
having spectroscopic metallicity estimates is 36\% (G04), 
hence B514 is in the low
metallicity tail of the distribution. However metal poor GCs are more abundant
in the outer regions of M31 (see Perrett et al. \cite{perr_cat}), therefore 
the metallicity of B514 is not unusual given its location.

Finally, comparing the measured indices with the predictions of the theoretical
models of Buzzoni (1989) and with the values observed in well studied galactic
globulars (Brodie \& Huchra \cite{bh0}) we find that B514 has 
$H_{\beta}$ and $Mgb$ values
typical of metal-poor clusters with age larger than 10 Gyr. Hence the available
photometric and spectroscopic data strongly indicates that B514 is a classical
old and metal-poor globular cluster.

%
\begin{table}
\caption{Line indices for the observed M31 clusters}  
\label{table:2}      
\centering                          
\begin{tabular}{l cccc}        
\hline\hline                 
Id. & $H_{\beta}$& Mgb &$H_{\beta}$ [BH] &Mgb [BH]\\    
(1) & (2)& (3) & (4) & (5)\\    
\hline                        
B158 & $0.084$    & $0.092$     & $0.085$ & $0.089$  \\
     & $\pm0.008$ & $\pm 0.008$ & $\pm 0.006$ &$\pm 0.007$\\
B225 & $0.053$    & $0.128 $    &$0.071$      &$0.130    $ \\      
     & $\pm0.008$ & $\pm 0.007$ & $\pm 0.003$ &$\pm 0.002$\\
B514 & $0.118$ & $0.030 $ &$-$ &$-$  \\ 
     & $\pm0.011$ & $\pm 0.010$ & $-$ &$-$\\
      
\hline                                   
\end{tabular}
{\begin{flushleft}{All the indices are expressed in magnitudes. Columns (2) and
(3)
report the estimates we obtained from our spectra, columns (4) and (5) report
the estimates published by Brodie \& Huchra (1990, BH). The indices are defined
as in BH.}
\end{flushleft}}
\end{table}
%
   \begin{figure}
   \centering
   \includegraphics[width=8cm]{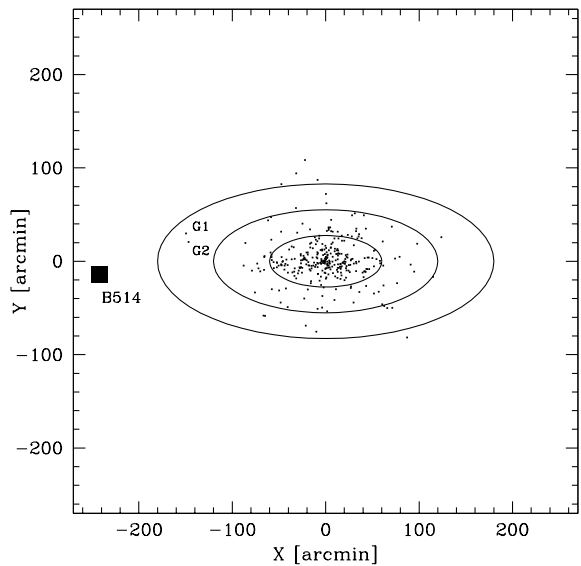}
      \caption{Distribution of M31 GCs (small dots)
      in the X, Y plane (distances 
      from the center of the galaxy projected along the major and 
      minor axis, respectively). Ellipses with the same axis ratio as the
      apparent disc of M31 and semi-major axis of $1\degr$, $2\degr$, and 
      $3\degr$ are over-plotted, for reference.  The
	most distant clusters (G1 and G2) and the new object B514 (filled
	square) are labeled.
               }
         \label{xy}
   \end{figure}
%
%
   \begin{figure}
   \centering
   \includegraphics[width=8cm]{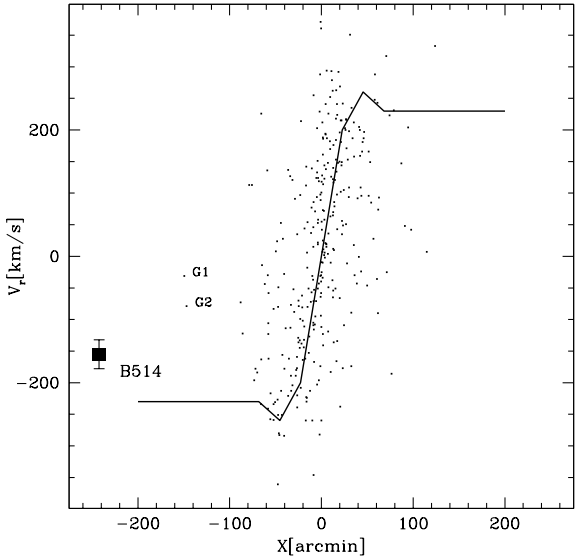}
      \caption{Radial velocity of the M31 globular clusters 
      (corrected for the systemic velocity of M31) vs. the projected
      distance along the major axis (X). 
      The HI rotation curve of the galaxy (from van den Berg \cite{sydbook})
      is plotted as a continuous line.  
              }
         \label{rotaz}
   \end{figure}
%

%
\begin{table}
\caption{Derived parameters for Bologna~514}             
\label{table:3}      
\centering                          
\begin{tabular}{l c }        
\hline\hline                 
 &  B514 \\    
\hline                        
$M_V$ & $-8.5 \pm 0.6$  \\      
$X$ & -242.40$\arcmin$ \\      
$Y$ & -14.16$\arcmin$ \\      
$R$ &  242.81$\arcmin$\\      
$R_p$ &  55.3 kpc \\      
$V_r$ & $-456 \pm 23$ km/s\\      
$[Fe/H]$& $-1.8 \pm 0.3$ \\      
\hline                                   
\end{tabular}
\end{table}

\subsection{Discussion}

The newly identified B514 cluster is placed in the realm of distances
inhabited  by the dwarf galaxy satellites of M31 (see Mateo \cite{mateo}) but
it doesn't lie in the surroundings of any of them, neither appears superposed
to the various low-surface-brightness structures and/or streams discovered in
recent surveys (see, for example, Ibata et al. \cite{iba01,iba04}, Zucker et
al. \cite{zuckNE,zuck9}, Merrett et al. \cite{merrett}, and references
therein).

On the other hand,  several authors have proposed that  the stellar population 
of the Andromeda galaxy (as sampled from Earth-centered line of sights) is
dominated by an extended disc component that provides the large majority of the
sampled stars out to $R_p\simeq 35$ kpc (Sarajedini \& Van Duyne \cite{sara},
Ferguson \& Johnson \cite{annette}, Worthey et al. \cite{worthey}). Moreover,
Morrison et al. (\cite{morrison}) showed that, at odds with the case of the
Milky Way, a significant fraction of M31 GCs share the positional and
kinematical properties of the dinamically-cold thin disc of the galaxy.
Fig.~\ref{xy} and ~\ref{rotaz}, above, shows that the position and M31-centric
velocity of B514 are fully compatible with the hypothesis that the cluster
belongs to the Morrison et al.'s thin disc clusters. If this will be confirmed
by further observations  it would imply a really huge size for the M31 disc,
with far reaching consequences on our ideas on the formation and evolution of
galactic discs.

Finally, it appears as a curious coincidence that the tree most remote M31
clusters known (e.g. B514, G1 and G2, see Fig.~\ref{xy} and ~\ref{rotaz}) are
all located in the proximity of the major axis and on the same side of the
galaxy (but B514 is more than $1^o$ farther than the other two; see also
Sect.~3.1 for dicussion about the metallicity). 
While their
M31-centric l.o.s. velocities differ by more than 100 km/s they are all
negative, e.g. in agreement with the sign of the rotation pattern of M31. Since
projection effects may have a significant r\^ole in producing the observed
differences in position and velocity, the possibility that the three systems
may be somehow related (as, for instance, by sharing similar orbits) cannot be
excluded at present and deserves further investigation.

\begin{acknowledgements}
We are grateful to the staff of the Loiano Observatory for their kind
assistance during the observing runs, in particular to Roberto Gualandi. We are
grateful to the INT WFC M31 survey team for their kind collaboration. This
publication makes use of data products from the Two Micron All Sky  Survey,
which is a joint project of the University of Massachusetts and the Infrared
Processing and Analysis Center/California Institute of Technology, funded by
the National Aeronautics and Space Administration and the National Science
Foundation. This research has made use of the SIMBAD database, operated at CDS,
Strasbourg, France and the NASA/IPAC Extragalactic Database (NED) which is
operated by the Jet Propulsion Laboratory, California Institute of Technology, 
under contract with the National Aeronautics and Space Administration. This
research has made use of NASA's Astrophysics Data System Abstract Service.

\end{acknowledgements}


\begin{thebibliography}{500}


\bibitem[1964]{BA} Baade, W., \& Arp, H. 1964, \apj, 139,
         1027

\bibitem[2000]{barm00} 
         Barmby, P., Huchra, J.P., Brodie J.P., et al. 2000, AJ, 119, 727
	 
\bibitem[2001]{barm01}
         Barmby, P., \& Huchra, J.P. 2001, AJ, 122, 2458	 

\bibitem[1980]{bat80}
         Battistini, P.L., Bonoli, F., Braccesi, A., Fusi Pecci, F., Malagnini, M.L.,
         Marano, B. 1980, A\&AS , 42, 357

\bibitem[1982]{bat82}
         Battistini, P.L., Bonoli, F., Buonanno, R., Corsi, C.E., 
	 Fusi Pecci, F. 1982, A\&A, 113, 39

\bibitem[1987]{bat87}
         Battistini, P.L., Bonoli, F., Braccesi, A., Federici, L., Fusi Pecci, F., Marano, B.,
	 Borngren F. 1987, A\&AS, 67, 447
	 
\bibitem[1993]{bat93} Battistini, P.L., Bonoli, 
         F., Casavecchia, M., Ciotti, L., Federici, L., \& Fusi Pecci, F. 
	 1993, \aap, 272, 77

\bibitem[1990]{bh0} Brodie, J.P., Huchra, J.P., 1990, ApJ, 362, 503	 

\bibitem[1989]{buz} Buzzoni, A., 1989, ApJS, 71, 817

\bibitem[1995]{buz95} Buzzoni, A., 1995, ApJS, 98, 69

\bibitem[1985]{ccsc85}
         Crampton, D., Cowley, A.P., Shade, D., Chayer, P. 1985, ApJ, 288, 494  

\bibitem[2003]{cutri} 
          Cutri et al. 2003, Explanatory Supplement to the 2MASS All Sky Data 
	  Release (see also:\\
	   {\tt http://www.ipac.caltech.edu/2mass/releases/allsky/ doc/explsup.html})

\bibitem[2004]{durrell}
         Durrell, P.R., Harris, W.E., \& Pritchet, C.J., 2004, AJ, 128, 260

\bibitem[2003]{evans}
         Evans, N.W., Wilkinson, M.I., Perrett, K.M., Bridges, T.J., 2003,
	 ApJ, 583, 752

\bibitem[1993]{luciana}
         Federici, L., Bonoli, F., Ciotti, L., et al., 1993, A\&A, 274, 87

\bibitem[2001]{annette}
         Ferguson, A.N.M, Johnson, R.A., 2001, ApJ, 599, L13	
	 
\bibitem[2004a]{silvia} 
        Galleti, S., Federici, L., Bellazzini, M., Fusi Pecci, F., Macrina, S., 
	2004a, A\&A, 416, 917 (G04)

\bibitem[2004b]{silviab} 
        Galleti, S., Cacciari, C., Federici, L., et al., 2004, in Stars in Galaxies,
	eds. M. Bellazzini, A. Buzzoni and S. Cassisi, Mem. SAIt, 75, 146
	
\bibitem[1996]{h96}
         Harris, W.E. 1996, AJ, 112, 1487  (see also:
	 {\tt http://physwww.physics.mcmaster.ca/\protect \\
	 $\sim$harris/mwgc.dat} for catalog update)

\bibitem[2001]{h01}
         Harris W.E. 2001, in {\it Star clusters}, Saas-Fee Advanced Course 28,
	 Lecture Notes 1998, eds. L. Labhardt and B. Biggeliwith the slit of $2\arcsec$, 
	 (Springer, Berlin), p.223 	 

\bibitem[1992]{h_book}
         Hodge, P.W. 1992, {\it The Andromeda galaxy}, Astrophys. and 
	 Space Science lib., vol. 176, Dordrecht, Kluwer
	 
\bibitem[1932]{hubble} 
         Hubble, E.P., 1932, ApJ, 76, 44	  
	 
\bibitem[1991]{bh}
         Huchra, J.P., Brodie, J.P., \& Kent, S.M. 1991, ApJ, 370, 495	 

\bibitem[2004]{hux}
         Huxor, A., Tanvir, N.R., Irwin, M.J., 2004, in Satellites and 
	 Tidal tails, eds. F. Prada, D. Martinez-Delgado and T. Mahoney,
	 (ASP, S. Francisco), ASP Conf. Ser., vol. 327, 118
	 
\bibitem[2001]{iba01} 
         Ibata, R., Irwin, M.J., Lewis, G., Ferguson, A.M.N., Tanvir, N.R., 
	 2001, Nature, 412, 49 

\bibitem[2004]{iba04}
         Ibata, R., Chapman, S., Ferguson, A.M.N., et al. 2004, MNRAS, 351, 117	

\bibitem[1998]{mateo}
         Mateo, M., 1998, ARA\&A, 36, 435

\bibitem[2005]{mcc}
         McConnachie, A.W., Irwin, M.J., Ferguson, A.M.N., Ibata R.A., 
	 Lewis, G.F., Tanvir, N., 2005, MNRAS, 356, 979

\bibitem[2003]{merrett}
         Merrett, H. R., Kuijken, K., Merrifield, M. R., 2003, MNRAS, 346, L62

\bibitem[2001]{g1}
         Meylan, G., Sarajedini, A., Jablonka, P., et al., 2001, AJ, 122, 830

\bibitem[1998]{Mochejska} Mochejska, B. J., Kaluzny, 
         J., Krockenberger, M., Sasselov, D.D., \& Stanek, K. Z. 1998, 
	 Acta Astron., 48, 455

\bibitem[2004]{morrison} 
         Morrison, H., Harding, P., Perrett, K.M., Hurley-Keller, D. 2004, 
	 ApJ, 603, 87 

\bibitem[2000]{vizier} 
         Ochsenbein, F., Bauer, P., Marcout, J. 2000, 
	 A\&AS, 143, 23

\bibitem[2003]{leda} 
         Paturel, G., Petit, C., Prugniel, Ph., Theureau, G., et al. 2003, 
	 A\&A, 412, 245

\bibitem[2002]{perr_cat} 
         Perrett, K.M., Bridges, T.J, Hanes, D.A., et al. 2002, 
	 AJ, 123, 2490

\bibitem[2001]{sara}
         Sarajedini, A., van Duyne, J., 2001 AJ, 122, 2444	 

\bibitem[1977]{sg77}
        Sargent, W.L.W., Kowal, C.T., Hartwick, F.D.A., van der Bergh S. 1977,
	 AJ, 82, 947

\bibitem[1995]{sha95}
        Sharov, A.F., Lyutyi, V.M., Esipov, V.F.  1995, PAstL, 21,240

\bibitem[1998]{trager}
         Trager, S.C., Worthey, G., Faber, S.M., Burstein, D., \& Gonz\'alez,
	 J.J 1998, ApJS, 116, 1 

\bibitem[1979]{tonry}
         Tonry, J., Davis, M., 1979, AJ, 84, 1511  

\bibitem[1967]{syd67}
         van den Bergh, S. 1967, AJ, 72, 70

\bibitem[1969]{syd69}
         van den Bergh, S. 1969, ApJS, 19, 145 

\bibitem[2000]{sydbook} 
        Van den Bergh, S., 2000, The galaxies of the Local Group,
	(Cambridge University Press, Cambridge)   
	
\bibitem[1962]{vet62}
        Vete$\check{s}$nik, M. 1962, Bull. Astron. Inst. Czechoslovakia, 
	13, 180
	
\bibitem[2004]{worthey}	 
         Worthey, G., Espana, A., MacArthur, L.A., Courteau, S., 2004, AJ,
	 submitted (astro-ph/0410454)         	

\bibitem[2004a]{zuckNE}
         Zucker, D.B., Kniazev, A.Y., Bell, E.F.,et al., 2004a, ApJ, 612, L117
	
\bibitem[2004b]{zuck9}
         Zucker, D.B., Kniazev, A.Y., Bell, E.F., et al., 2004b, ApJ, 612, L121
	 

\end{thebibliography}
\end{document}